\documentclass[11pt,twoside]{article}
\usepackage{asp2010}

\resetcounters

\begin{document}

\allowtitlefootnote

\title{Be star outbursts: transport of angular momentum by waves}
\author{C. Neiner,$^1$ S. Mathis,$^2$ H. Saio,$^3$ and U. Lee$^3$
\affil{$^1$LESIA, Observatoire de Paris, CNRS UMR 8109, UPMC, Universit\'e Paris Diderot; 5 place Jules Janssen, 92190 Meudon, France\\
$^2$Laboratoire AIM Paris-Saclay, CEA/DSM - CNRS - Universit\'e Paris
Diderot, IRFU/Service d'Astrophysique, CEA-Saclay, 91191 Gif-sur-Yvette Cedex,
France\\
$^3$Astronomical Institute, Graduate School of Science, Tohoku University, Sendai, 980-8578, Japan}}

\begin{abstract}
The Be phenomenon, that is the ejection of matter from Be stars into a
circumstellar disk, has been a long lasting mystery. In the last few years, the
CoRoT (Convection, Rotation and planetary Transits) satellite brought clear
evidence that Be outbursts are directly correlated with pulsations. We found
that it may be the transport of angular momentum by waves or pulsation modes
that brings the already rapid stellar rotation to its critical value at the
surface, and allows the star to eject material. The recent discovery of
stochastically excited gravito-inertial modes by CoRoT in a hot Be star
strengthens this scenario. We present the CoRoT observations and modeling of
several Be stars and describe the new picture of the Be phenomenon which arose
from these results. 
\end{abstract}

\section{Be stars and the Be phenomenon}

Be stars are Main Sequence or slightly evolved non-supergiant, late-O, B, or
early-A stars that show or have shown emission in at least one Balmer line
\citep{collins1987}. Emission can also appear in other lines of the spectrum as
well as in the continuum, particularly as an excess of light in the infrared
domain. The emission is due to the presence of a circumstellar disk, fed by
episodic ejections of material from the surface of the star to a Keplerian
orbit. This is called the ``Be phenomenon''. However, how these outbursts
producing the disk can occur was not understood until now.

There are $\sim$2000 Be stars known as of today and listed in the BeSS (Be Star
Spectra) database of Be stars \citep{neiner_BeSS}. Be stars represent about 20\%
of all B-type stars in our galaxy. Only B stars of sufficiently high rotational
velocity at the Zero Age Main Sequence (ZAMS) can become Be stars
\citep{martayan2007}, and this velocity depends on the metallicity of the
protostellar environment. In addition, there seems to be a strong dependence of
the proportion of Be stars compared to B stars as a function of spectral type.
Indeed, the Be phenomenon is mostly observed around the sub-types B1-B2
\citep[see, e.g., Fig.~1 of][]{balona2000}.

B2 is also the spectral type at which pulsations of both $\beta$\,Cep and Slowly
Pulsating B (SPB)  types can occur at the same time. Light and line-profile
short-term variability (of the order of one day) due to non-radial pulsations is
indeed detected from the ground or with Hipparcos in almost all (86\%) early-Be
stars, in 40\% of mid-types (B4-5e), and in only 18\% of late-Be stars according
to \cite{hubert1998}. Thanks to high-precision photometric space-based missions,
we actually find that all Be stars pulsate, whatever their spectral type, but
with a lower amplitude for cooler stars. Short-term variability also occurs
owing to rotational modulation caused for example by surface spots.

Variability on other time scales is also common. Cyclic variations of the order
of months or years are associated with the wind, which is variable and stronger
than for B stars. In addition, variations in the disk emission line profiles with time-scales of a few years to decades are observed: first, the
intensity of emission lines slowly decreases as the disk dissipates in the
interstellar medium; second, a denser zone can exist in the disk, which slowly
precesses and produces global one-armed disk oscillation. Finally, abrupt
emission increases are due to short-lived (days, tens of days), sometimes
recurrent, or/and long-lived (months) outbursts. Depending on the frequency of
outbursts and on the speed of disk dissipation, the disk can sometimes
completely disappear and reappear following a new outburst. For a complete
review of Be stars, we refer the reader to \cite{porter2003}.

Be stars are known to be very fast rotators. This leads to many rotational
effects, in particular Be stars are flattened by the centrifugal force. However,
their velocities are not high enough to reach the critical limit at which the
centrifugal force compensates gravity at the equator. Indeed, galactic Be stars
rotate on average at 88\% of the critical angular velocity  \citep{fremat2005}.
Thus, at least in most cases, while rotation is certainly an important
ingredient in igniting outbursts, it cannot  by itself explain the ejection of
matter from the star that leads to the formation of the decretion disk. Another
mechanism is required to provide the additional angular momentum at the surface
needed to eject matter. 

Several explanations have been put forward to provide this additional angular
momentum. \cite{oudmaijer2010} showed that about 30\% of Be stars are in binary
systems with a close companion, which is similar to the binary rate of B stars.
The binary companion certainly plays a very important role for those 30\% of Be
stars (e.g., in Be/X-ray binaries). For single stars, results from the MiMeS
project \citep[e.g.][]{neiner2011} showed that the magnetic field of Be stars,
if it is present, is weak, and that it only influences possible co-rotating
clouds close to the stellar surface and not the Keplerian Be disk
\citep{neiner_omeori2012}. Therefore pulsations appear to be the most likely
explanation, in addition to the rapid rotation, of the Be phenomenon. 

\section{Correlation between pulsations and outbursts in HD\,49330}
\label{hd49330}

CoRoT allowed us to observe the hot (B0IVe) Be star HD\,49330 during an outburst
\citep{huat2009}. By completing a photometric analysis of the precise CoRoT data
of that star acquired over $\sim$136 days, we were able to detect over 300
different frequencies attributable to variations, including at least thirty
independent ones. These include  high frequencies as well as several groups of
low frequencies, which are typical of the pressure (p) and gravity (g) pulsation
modes observed in $\beta$ Cep and SPB stars, respectively. Some of the
frequencies have also been detected in photospheric line profile variations with
the help of simultaneous spectroscopic observations \citep{floquet2009}.

Thanks to these CoRoT data, we have discovered a correlation between both
amplitude changes and the presence/absence of certain frequencies of pulsation,
with the different phases of an outburst: (i) the amplitude of the main 
p-mode frequencies decreases before and during the outburst and increases
again after it; (ii) several groups of g-mode frequencies
appear just before the outburst, their amplitude reaches a maximum during the
outburst, and they disappear as soon as it has finished. These
frequencies appear to have complex structures, which could represent pulsation
modes with a short lifetime.

The detailed characterization of the pulsation modes and fundamental stellar
parameters of HD\,49330, and the presence of both p-modes and g-modes, provide strong
constraints on its seismic modeling. We tried to model HD\,49330 with
$\kappa$-driven pulsations using the Tohoku oscillation code
\citep{saiocox1980,lee1995} that accounts for the combined action of Coriolis
and centrifugal accelerations on stellar pulsations as needed for Be star
modeling. However, whatever models of pulsations and stellar structure we used,
we find that p-modes and g-modes cannot be excited at the same time by the
$\kappa$ mechanism in HD\,49330 using stellar parameters determined
spectroscopically. 

\section{Discovery of stochastically excited gravito-inertial modes in HD\,51452}

HD\,51452 is a hot Be (B0IVe) star observed with CoRoT, which has a rotation
frequency $f_{\rm rot} \sim 1.22$ c~d$^{-1}$. For such a hot star, the $\kappa$
mechanism can create p modes of pulsations or possibly low-order g modes with
frequencies above $\sim$1.5 c~d$^{-1}$. Nevertheless g-modes are detected in the
CoRoT light curve of HD\,51452 with frequencies below $\sim$1.5 c~d$^{-1}$. In
addition, a multiplet of frequencies is detected, with its main peak in the
domain below $\sim$1.5 c~d$^{-1}$, with a frequency spacing of about 0.6
c~d$^{-1}$ \citep[see][]{neiner_51452}. These frequencies below $\sim$1.5
c~d$^{-1}$ cannot be explained with the $\kappa$ mechanism, even when taking
stellar flattening into account. They could, however, be stochastic
gravito-inertial (gi) modes. In particular sub-inertial gi-modes would be
below $\sim$2.44 c~d$^{-1}$.

There are two types of gi-mode: (1) those usually called g-modes, which are
gravity modes modified by the Coriolis acceleration, and which show a regular
pattern in period when they are asymptotic \citep{lee1997,ballot2010}, and (2)
rotational (r) modes, which are mainly driven by the Coriolis acceleration, are
sub-inertial and show regular patterns in frequency
\citep{provost1981,saio1982,lee2006}. Since we observed at least one multiplet
in frequency in the sub-inertial domain for HD\,51452, those peaks may be
interpreted as r-modes. Since no specific frequency or period spacing is found 
for the other frequency peaks, they could be any type of gi-mode.

Convective regions, such as the convective core and the sub-surface convection
zone in massive stars, are indeed able to stochastically excite oscillation
modes \citep{cantiello2009,belkacem2010} and particularly g-modes
\citep{samadi2010,shiode2013}. The latter become gi-modes in fast rotators such
as HD\,51452 because of the action of the Coriolis acceleration \citep[see
e.g.][]{lee1997,DR2000,Mathis2009,ballot2010}. This excitation is also observed
in realistic numerical simulations of convective cores surrounded by a stably
stratified radiative envelope \citep[see][]{browning2004}. Gravito-inertial
waves are excited through their couplings with volumetric turbulence in
convective regions (where pure g-modes in a slowly rotating star are evanescent,
while gi-modes in a rapidly rotating star become inertial) and by the impact of
structured turbulent plumes at the interfaces between convective and radiative
regions.

\cite{samadi2010} examined the stochastic excitation of gravity modes by
turbulent convection in massive stars. They found that the excitation of low
$n$-order g-modes occurs in the core while the asymptotic g-modes are mostly
excited in the outer convective zone. The mode amplitudes that they deduced,
however, are well below the detection threshold of the CoRoT satellite for a
massive star. On the contrary, recent work by \cite{shiode2013} showed
analytically that taking stochastic excitation by penetration into account
produces g-modes of detectable amplitude. In both works however, no rotation is
considered. In addition, three-dimensional (3D) simulations of a convective core
in a A star \citep{browning2004} or B star (Augustson et al., in preparation)
showed efficient stochastic excitation of g-waves.

HD\,51452 is a Be star, rotating close to its breakup velocity. Therefore, the
calculation of the excitation of gi-modes (including r-modes) in this star 
requires the study of the influence of very fast rotation. This application can
be derived from the work by \cite{belkacem2009} and is the study of a
forthcoming paper (Mathis et al. 2013). The detection of gi-modes in
HD\,51452 presented in \cite{neiner_51452}, however, already suggests that fast
rotation enhances the amplitude of gi-modes and thus r-modes.

The fact that HD\,51452 is a very hot Be star excludes the possibility that the
observed gi-modes are excited by the $\kappa$-mechanism. In view of these
results, however, it might be necessary to reconsider our interpretation of
several other rapidly rotating B or Be stars, for which the $\kappa$-mechanism
seemed like an obvious excitation mechanism but for which stochastic excitation
might also be at work. In particular the g-modes observed in HD\,49330 (see
Sect.~\ref{hd49330}) could be stochastically excited, which would explain why
the seismic models with $\kappa$-driven modes did not work.

\section{Transport of angular momentum by waves in Be stars}

Considering the arguments presented above for HD\,51452, the results of our
modeling of HD\,49330 with the $\kappa$-mechanism, as well as the fact that the
power spectrum of that star shows broad frequency groups rather than sharp
frequency peaks around 1-2 c~d$^{-1}$ \citep{huat2009}, we propose that
HD\,49330 hosts stochastic g-waves. 

During the quiet phase, stochastic gi-waves can be excited in the convective
core. If they are sub-inertial, as observed in the Be star HD\,51452, these
waves transport more angular momentum to the subsurface layers than
$\kappa$-driven modes because their frequency is lower. The net deposit of
angular momentum indeed depends on the thermal dissipation, i.e., it is 
proportional to $1/f^4$ for a given rotation \citep{zahn1997, Mathis2009}. When
enough angular momentum has accumulated in the outer layers of the star, these
layers get unstable and could emit transient g-waves, which we detect. Possibly,
the g-waves excited in the core may then also become visible. The surface layers
reach the critical velocity, in particular at the equator where the rotation was
already the closest to critical. The destabilisation of the surface layers thus
ignites the outburst and breaks the cavity in which the p-modes were
propagating. This explains both the disappearance of the p-modes during the
precursor and outburst phases of HD\,49330 and the ejection of material from the
surface into the disk, i.e., the occurrence of the outburst, as observed by
\cite{huat2009} with CoRoT. Relaxation then occurs, recreating the cavity and
letting the p-modes reappear while the transient g-waves disappear. Each time
stochastic g-waves from the core accumulate enough angular momentum in the outer
layers of the star, this outburst phenomenon will occur again.

The idea that non-radial oscillations excited in massive stars are able to
efficiently transport angular momentum and to allow the surface of a Be star to
reach its critical velocity has already been proposed by \cite{ando1986},
\cite{lee_saio}, and \cite{lee2006}. However, in these previous works, gi-waves
were excited by the $\kappa$-mechanism. In the case of HD\,49330, we propose
that gi-waves are stochastically excited in turbulent convective regions.
\cite{pantillon2007} and \cite{rogers2012} demonstrated that stochastic gi-waves
are able to transport angular momentum in the same way as $\kappa$-driven waves.
The type of excitation does not influence the transport. It depends mostly on
the amplitude and frequency of the mode \citep{zahn1997, Mathis2009}. The higher
the amplitude and the lower the frequency, the more transport there is. 

The Tohoku models of HD\,49330 indeed show that the rate of local angular
momentum change due to pulsational transportation integrated over the mean
sphere increases drastically in the few percents of the stellar radius just
below the surface, i.e., there is a strong net deposit of angular momentum in
the surface layers, because this is where the pulsations are damped. This
confirms that pulsations increase angular momentum in the outermost layers of
the star. 

The Tohoku pulsation code is currently being modified to include stochastically
excited pulsations. This new version will be used to test our
scenario (Neiner, Saio et al., in preparation).

\section{Conclusion}

All Be stars observed with CoRoT pulsate, whatever their spectral type. A
scenario similar to the one observed in HD\,49330 could thus occur in all other
Be stars, thus providing an explanation of the long-lasting mystery of the
origin of Be outbursts and disks, for single Be stars.

Stochastic waves are certainly excited in the convective core of any massive
star, but with amplitudes that may be undetectable (even with space-based
facilities) if the star rotates slowly. From the observations of stochastically
excited gi-modes in HD\,51452 and the results of the seismic modeling of HD\,49330, we
however propose that rotation enhances the amplitude of stochastic
gi-waves/modes. Since these waves are of low frequencies, they transport more
angular momentum than other waves/modes of higher frequency for a given
amplitude. Therefore we conclude that a hot slowly rotating pulsating B star
will only excite detectable $\kappa$-driven p modes, i.e., it will be a
$\beta$\,Cep star, but if it rotates fast, it will also excite stochastic
gi-waves/modes with a larger amplitude than the $\beta$\,Cep star and it will
become a hot Be star. A cool slowly rotating pulsating B star will excite
$\kappa$-driven g-modes, i.e., it will be a SPB star, but if it rotates fast, it
will also excite stochastic gi-waves/modes with a larger amplitude than the SPB
star and it will become a cool Be star.

We conclude that single Be stars bring their surface layer above the critical
velocity thanks to three ingredients: (1) rapid rotation itself, (2) the
transport of angular momentum by pulsations (of all types, but mostly prograde
g-modes), and (3) the enhancement of the amplitude of stochastic gi-waves thanks
to the rapid rotation, which allows more transport of angular momentum. The fact
that a B star becomes a Be star therefore depends strongly on its rotation rate,
pulsations, and convection (in competition with its stellar wind).

\acknowledgements The CoRoT space mission, launched on December 27th 2006, has
been developed and is operated by Centre National d'Etudes Spatiales (CNES),
with the contribution of Austria, Belgium, Brazil, the European Space Agency
(RSSD and Science Program), Germany and Spain. CN acknowledges fundings from
SIROCO (Seismology, Rotation and Convection with the CoRoT Satellite) ANR
(Agence Nationale de la Recherche) project, CNES and PNPS (Programme National de
Physique Stellaire).

\bibliography{references_Neiner}

\begin{thebibliography}{}
\expandafter\ifx\csname natexlab\endcsname\relax\def\natexlab#1{#1}\fi
\expandafter\ifx\csname url\endcsname\relax
  \def\url#1{\texttt{#1}}\fi
\expandafter\ifx\csname urlprefix\endcsname\relax\def\urlprefix{URL }\fi
\providecommand{\eprint}[2][]{\url{#2}}

\bibitem[{{Ando}(1986)}]{ando1986}
{Ando}, H. 1986, \aap, 163, 97

\bibitem[{{Ballot} et~al.(2010){Ballot}, {Ligni{\`e}res}, {Reese}, \&
  {Rieutord}}]{ballot2010}
{Ballot}, J., {Ligni{\`e}res}, F., {Reese}, D.~R., \& {Rieutord}, M. 2010,
  \aap, 518, A30

\bibitem[{{Balona}(2000)}]{balona2000}
{Balona}, L.~A. 2000, in IAU Colloq. 175: The Be Phenomenon in Early-Type
  Stars, edited by M.~A. {Smith}, H.~F. {Henrichs}, \& J.~{Fabregat}, vol. 214
  of ASPC, 1

\bibitem[{{Belkacem} et~al.(2010){Belkacem}, {Dupret}, \&
  {Noels}}]{belkacem2010}
{Belkacem}, K., {Dupret}, M.~A., \& {Noels}, A. 2010, \aap, 510, A6

\bibitem[{{Belkacem} et~al.(2009){Belkacem}, {Mathis}, {Goupil}, \&
  {Samadi}}]{belkacem2009}
{Belkacem}, K., {Mathis}, S., {Goupil}, M.~J., \& {Samadi}, R. 2009, \aap, 508,
  345

\bibitem[{{Browning} et~al.(2004){Browning}, {Brun}, \&
  {Toomre}}]{browning2004}
{Browning}, M.~K., {Brun}, A.~S., \& {Toomre}, J. 2004, \apj, 601, 512

\bibitem[{{Cantiello} et~al.(2009){Cantiello}, {Langer}, {Brott}, {de Koter},
  {Shore}, {Vink}, {Voegler}, {Lennon}, \& {Yoon}}]{cantiello2009}
{Cantiello}, M., {Langer}, N., {Brott}, I., et al. 2009, \aap,
  499, 279

\bibitem[{{Collins}(1987)}]{collins1987}
{Collins}, G.~W., II 1987, in IAU Colloq. 92: Physics of Be Stars, edited by
  A.~{Slettebak}, \& T.~P. {Snow}, 3

\bibitem[{{Dintrans} \& {Rieutord}(2000)}]{DR2000}
{Dintrans}, B., \& {Rieutord}, M. 2000, \aap, 354, 86

\bibitem[{{Floquet} et~al.(2009){Floquet}, {Hubert}, {Huat}, {Fr{\'e}mat},
  {Janot-Pacheco}, {Guti{\'e}rrez-Soto}, {Neiner}, {de Batz}, {Leroy},
  {Poretti}, {Amado}, {Catala}, {Rainer}, {Diaz}, {Uytterhoeven}, {Andrade},
  {Diago}, {Emilio}, {Espinosa Lara}, {Fabregat}, {Martayan}, {Semaan}, \&
  {Suso}}]{floquet2009}
{Floquet}, M., {Hubert}, A.-M., {Huat}, A.-L., {Fr{\'e}mat}, Y.,
  {Janot-Pacheco}, E., {Guti{\'e}rrez-Soto}, J., {Neiner}, C., {de Batz}, B.,
  {Leroy}, B., {Poretti}, E., {Amado}, P., {Catala}, C., {Rainer}, M., {Diaz},
  D., {Uytterhoeven}, K., {Andrade}, L., {Diago}, P.~D., {Emilio}, M.,
  {Espinosa Lara}, F., {Fabregat}, J., {Martayan}, C., {Semaan}, T., \& {Suso},
  J.et al. 2009, \aap, 506, 103

\bibitem[{{Fr{\'e}mat} et~al.(2005){Fr{\'e}mat}, {Zorec}, {Hubert}, \&
  {Floquet}}]{fremat2005}
{Fr{\'e}mat}, Y., {Zorec}, J., {Hubert}, A.-M., \& {Floquet}, M. 2005, \aap,
  440, 305

\bibitem[{{Huat} et~al.(2009){Huat}, {Hubert}, {Baudin}, {Floquet}, {Neiner},
  {Fr{\'e}mat}, {Guti{\'e}rrez-Soto}, {Andrade}, {de Batz}, {Diago}, {Emilio},
  {Espinosa Lara}, {Fabregat}, {Janot-Pacheco}, {Leroy}, {Martayan}, {Semaan},
  {Suso}, {Auvergne}, {Catala}, {Michel}, \& {Samadi}}]{huat2009}
{Huat}, A.-L., {Hubert}, A.-M., {Baudin}, F., et al. 2009, \aap, 506,
  95

\bibitem[{{Hubert} \& {Floquet}(1998)}]{hubert1998}
{Hubert}, A.~M., \& {Floquet}, M. 1998, \aap, 335, 565

\bibitem[{{Lee}(2006)}]{lee2006}
{Lee}, U. 2006, \mnras, 365, 677

\bibitem[{{Lee} \& {Baraffe}(1995)}]{lee1995}
{Lee}, U., \& {Baraffe}, I. 1995, \aap, 301, 419

\bibitem[{{Lee} \& {Saio}(1993)}]{lee_saio}
{Lee}, U., \& {Saio}, H. 1993, \mnras, 261, 415

\bibitem[{{Lee} \& {Saio}(1997)}]{lee1997}
--- 1997, \apj, 491, 839

\bibitem[{{Martayan} et~al.(2007){Martayan}, {Fr{\'e}mat}, {Hubert}, {Floquet},
  {Zorec}, \& {Neiner}}]{martayan2007}
{Martayan}, C., {Fr{\'e}mat}, Y., {Hubert}, A.-M., {Floquet}, M., {Zorec}, J.,
  \& {Neiner}, C. 2007, \aap, 462, 683

\bibitem[{{Mathis}(2009)}]{Mathis2009}
{Mathis}, S. 2009, \aap, 506, 811

\bibitem[{{Neiner} et~al.(2011{\natexlab{a}}){Neiner}, {Alecian}, \&
  {Mathis}}]{neiner2011}
{Neiner}, C., {Alecian}, E., \& {Mathis}, S. 2011{\natexlab{a}}, in SF2A-2011:
  Proceedings of the Annual meeting of the French Society of Astronomy and
  Astrophysics, 509

\bibitem[{{Neiner} et~al.(2011{\natexlab{b}}){Neiner}, {de Batz}, {Cochard},
  {Floquet}, {Mekkas}, \& {Desnoux}}]{neiner_BeSS}
{Neiner}, C., {de Batz}, B., {Cochard}, F., {Floquet}, M., {Mekkas}, A., \&
  {Desnoux}, V. 2011{\natexlab{b}}, \aj, 142, 149

\bibitem[{{Neiner} et~al.(2012{\natexlab{a}}){Neiner}, {Floquet}, {Samadi},
  {Espinosa Lara}, {Fr{\'e}mat}, {Mathis}, {Leroy}, {de Batz}, {Rainer},
  {Poretti}, {Mathias}, {Guarro Fl{\'o}}, {Buil}, {Ribeiro}, {Alecian},
  {Andrade}, {Briquet}, {Diago}, {Emilio}, {Fabregat}, {Guti{\'e}rrez-Soto},
  {Hubert}, {Janot-Pacheco}, {Martayan}, {Semaan}, {Suso}, \&
  {Zorec}}]{neiner_51452}
{Neiner}, C., {Floquet}, M., {Samadi}, R., et al. 2012{\natexlab{a}},
  \aap, 546, A47

\bibitem[{{Neiner} et~al.(2012{\natexlab{b}}){Neiner}, {Grunhut}, {Petit},
  {ud-Doula}, {Wade}, {Landstreet}, {de Batz}, {Cochard}, {Guti{\'e}rrez-Soto},
  \& {Huat}}]{neiner_omeori2012}
{Neiner}, C., {Grunhut}, J.~H., {Petit}, V., et al. 2012{\natexlab{b}}, \mnras, 426, 2738

\bibitem[{{Oudmaijer} \& {Parr}(2010)}]{oudmaijer2010}
{Oudmaijer}, R.~D., \& {Parr}, A.~M. 2010, \mnras, 405, 2439

\bibitem[{{Pantillon} et~al.(2007){Pantillon}, {Talon}, \&
  {Charbonnel}}]{pantillon2007}
{Pantillon}, F.~P., {Talon}, S., \& {Charbonnel}, C. 2007, \aap, 474, 155

\bibitem[{{Porter} \& {Rivinius}(2003)}]{porter2003}
{Porter}, J.~M., \& {Rivinius}, T. 2003, \pasp, 115, 1153

\bibitem[{{Provost} et~al.(1981){Provost}, {Berthomieu}, \&
  {Rocca}}]{provost1981}
{Provost}, J., {Berthomieu}, G., \& {Rocca}, A. 1981, \aap, 94, 126

\bibitem[{{Rogers} et~al.(2012){Rogers}, {Lin}, \& {Lau}}]{rogers2012}
{Rogers}, T.~M., {Lin}, D.~N.~C., \& {Lau}, H.~H.~B. 2012, \apjl, 758, L6.
  \eprint{1209.2435}

\bibitem[{{Saio}(1982)}]{saio1982}
{Saio}, H. 1982, \apj, 256, 717

\bibitem[{{Saio} \& {Cox}(1980)}]{saiocox1980}
{Saio}, H., \& {Cox}, J.~P. 1980, \apj, 236, 549

\bibitem[{{Samadi} et~al.(2010){Samadi}, {Belkacem}, {Goupil}, {Dupret},
  {Brun}, \& {Noels}}]{samadi2010}
{Samadi}, R., {Belkacem}, K., {Goupil}, M.~J., {Dupret}, M.-A., {Brun}, A.~S.,
  \& {Noels}, A. 2010, \apss, 328, 253

\bibitem[{{Shiode} et~al.(2013){Shiode}, {Quataert}, {Cantiello}, \&
  {Bildsten}}]{shiode2013}
{Shiode}, J.~H., {Quataert}, E., {Cantiello}, M., \& {Bildsten}, L. 2013,
  \mnras, 690. \eprint{1210.5525}

\bibitem[{{Zahn} et~al.(1997){Zahn}, {Talon}, \& {Matias}}]{zahn1997}
{Zahn}, J.-P., {Talon}, S., \& {Matias}, J. 1997, \aap, 322, 320

\end{thebibliography}

\end{document}